\documentstyle[11pt,paspconf,epsf]{article}

\begin{document}

\title{How Galaxies Don't Form \\
{\small The Effective Force Law in Disk Galaxies}}
\author{Stacy McGaugh}
\affil{Department of Astronomy, University of Maryland, USA}

\begin{abstract}
I review current theories of disk galaxy formation.
All fail in the light of recent data for low surface brightness galaxies.
The difficulty stems from the presence of a unique acceleration scale
in the data.
\end{abstract}


\begin{quote}
\raggedleft
All we know so far is what doesn't work. \\
--- Richard Feynman
\end{quote}

\section{Introduction}

Galaxy formation is an intriguing but difficult subject.
Substantial progress has been made over the last few years, culminating in
a rough consensus on a ``standard'' picture of galaxy formation (e.g.,
Mo et al.\ 1998).  In spite of my own modest contributions to this
(e.g., Mo et al.\ 1994), it is my contention that this picture is wrong.

Empirically, there are a number of striking regularities in the dynamical
data for disk galaxies.  These enable us to describe
the effective force law.  It is remarkably uniform, and depends only on
the luminous mass.

\section{Galaxy Formation}

The general picture we have of disk galaxy formation is of cool gas
settling into a disk in a potential well dominated by dark matter.  The
dark matter is assumed to reside in a dissipationless, effectively spherical
distribution, and to interact with the potentially luminous (``baryonic'')
matter only gravitationally.  To make predictions from this general picture,
we need to specify something about both the halos and the baryons.
I would like to emphasize the importance of making {\it a priori\/}
predictions.  If we simply tune our favorite model to fit to the data,
we will certainly fit the data.

At minimum, two parameters are required to describe halos, such as a
characteristic mass ${\cal M}_h$ and size $R_h$.  Further parameters describing
axial and kinematic isotropy, etc., could be invoked, but these distract from
the basic point and are not fundamental.  For the baryons, many parameters
might in principle come into play: initial angular momentum and gas
temperature, conversion of gas into stars and feedback into the ISM,
and so on.  It would be nice to study a population of galaxies for which
the effects of the baryons were minimized.
Low surface brightness (LSB) galaxies, with central surface brightnesses
$\mu_0 > 23\>B\,{\rm mag.}\,{\rm arcsec}^{-2}$, turn out to be just
such objects.  They occupy a different region of
parameter space than do high surface brightness
(HSB) spirals, so they provide genuinely
new tests of ideas contrived to explain observations of HSB disks.

\subsection{Density Begets Density}

Since I am going to contradict some widely held notions,
it seems only fair to start by trashing my own favorite idea.
Many of the properties of LSB galaxies (McGaugh 1992)
suggested to me that they were basically stretched
out, lower density versions of HSB galaxies.  They also
appear to be somewhat younger than HSB galaxies, suggesting a later collapse
epoch (McGaugh \& Bothun 1994).

It seems natural that the properties of dark halos might
dictate the properties of the luminous galaxies they contain.  In
addition to a distribution of masses ${\cal M}_h$ which gives rise to the
luminosity function, there could also be a distribution of scale sizes
$R_h$ which, at a given mass, gives rise to the distribution of disk
scale lengths (or equivalently, the surface brightness distribution:
McGaugh 1996).  The mapping from (${\cal M}_h, R_h$) to ($L, h$) need not
be simple, but assuming it is forms an obvious starting point.
In this picture (which I refer to as `density begets density,' or DD),
HSB disks arise from large density
fluctuations $\delta$, and LSB disks from low $\delta$ (but not necessarily
low mass).

DD makes several predictions beyond the data which motivated it
(McGaugh \& de Blok 1998a).
One is that since HSB and LSB galaxies represent populations which arose from
different characteristic $\delta$, there should be a shift between the
correlation function of the two.  This prediction
was confirmed (Mo et al.\ 1994).

In DD, LSB galaxies and their halos are stretched out versions of HSB systems.
Two predictions follow about their dynamics.
They should have slowly rising rotation curves (which is true),
and they should deviate systematically from the Tully-Fisher (TF) relation.

The expected departure from the TF relation goes in the sense
that LSB galaxies should rotate slowly for their luminosity.  This follows
simply because $R_h$ is large for a given ${\cal M}_h$, and
$V^2 = G {\cal M}/R$.  There is some freedom to
tune the amount of the shift, so long as
it is systematic with surface brightness.  It would not
fit nicely into this picture if there were zero shift.

\begin{figure}[t]
\plotone{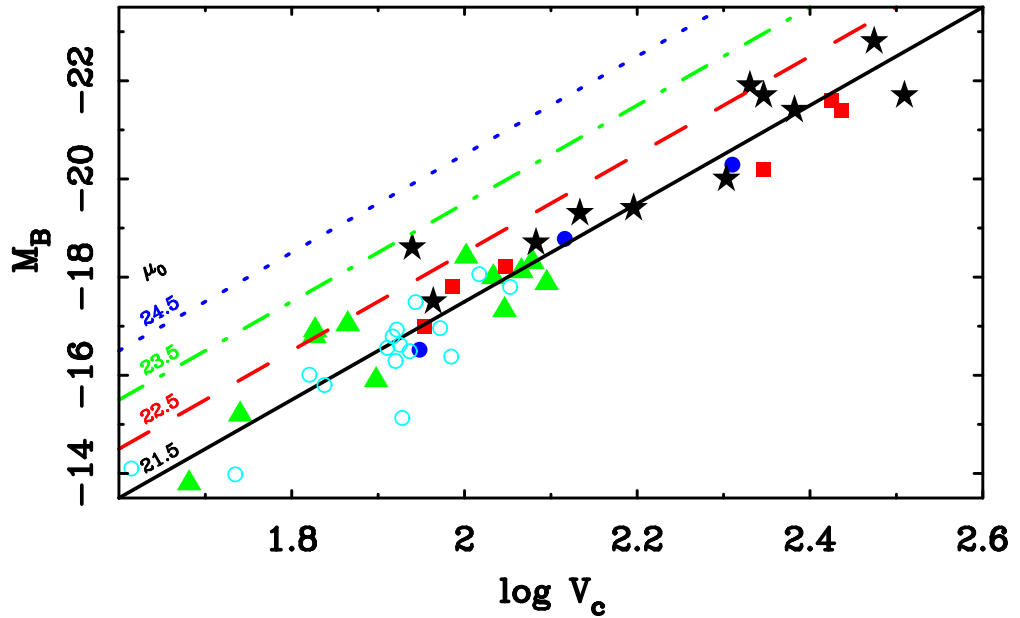}
\caption{The luminosity--rotation velocity (Tully-Fisher) relation.
Data are shown for galaxies with $V_c$ measured from the flat part of
resolved, extended rotation curves (solid symbols; de Blok et al.\ 1996) and
for some with recent high quality velocity widths (open symbols;
Matthews et al.\ 1998).  Symbols distinguish different bins of central surface
brightness --- stars: $\mu_0 < 22$; squares: $22 < \mu_0 < 23$; triangles:
$23 < \mu_0 < 24$; filled circles: $\mu_0 > 24$.  Galaxies
with open symbols predominantly have $\mu_0 > 23$.  The data do not distinguish
themselves by surface brightness in this diagram.  The lines are not fits to
the data, but have the ``virial'' slope of $-10$.  They are labeled by
a corresponding central surface brightness illustrating the systematic shift
that is expected from simple arguments (see text).
}
\label{TFSB}
\end{figure}

This is exactly what is observed (Figure 1).  Regardless of surface brightness,
LSB galaxies fall on the same\footnote{This result has been independently
obtained by Zwaan et al.\ (1995), Sprayberry et al.\ (1995),
Hoffman et al.\ (1996), and Tully \& Verheijen (1997).
It has been disputed by Matthews et al.\ (1998), who suggest that some
faint LSB galaxies fall systematically below the TF relation.
This is a result of comparing to a TF relation with a shallow slope
determined from a fit to bright galaxies.  The {\it data\/} of Matthews
et al.\ are consistent with other data (Fig.\ 1).  Indeed, they add
weight to the faint end which is much needed for constraining the slope.}
TF relation with the same normalization
as HSB galaxies.  The shift I expected
is illustrated by the lines in Figure 1, and clearly does not occur.

\subsection{Same Halo}

As an alternative to DD, I have investigated
the class of theories which I generically label `same
halo' (SH; McGaugh \& de Blok 1998a).  The basic idea here is that there is
{\it no\/} distribution in $R_h$. 
Simulations of halo formation (e.g., Navarro et al.\ 1997; hereafter NFW)
indicate a strong correlation between halo parameters, consistent with
this picture.  At a given mass, the halo is the same
regardless of the scale length of the optical galaxy.
This yields the desired TF relation, by construction.

Since there is now no distribution in $R_h$, we must
invoke some other mechanism to give the observed distribution of optical
surface brightnesses.  This is usually assumed to follow from the initial
angular momentum.  In terms of Peebles's spin parameter $\lambda$, the scale
length of the luminous disk is
$h \approx \lambda R_h$.  The precise equation can be more complicated
(Dalcanton et al.\ 1997\footnote{Given the emphasis placed on {\it a
priori\/} predictions, it is worth noting that the original version of
the scenario described by Dalcanton et al.\ (astro-ph/9503093)
predicted a {\it shift\/} in the TF relation with surface brightness.
The observed lack of such a shift was reported at about this time.
Zero shift is `predicted' in the published version.}; Mo et al.\ 1998), but
this encapsulates the basic idea.  Baryons in a halo with low initial spin
collapse a long way before rotational support is achieved, forming
an HSB galaxy with a short scale length.  A high spin halo of the same mass
forms an LSB galaxy with a much larger disk scale length.  This makes
the rather dubious assumption that there is no interaction between
disk and halo which can transfer angular momentum between the two.

The surface brightness distribution is now determined by the initial
distribution of $\lambda$ rather than $\delta$, causing a different
problem to arise.  In fixing the failings of DD with regards to the
TF relation, we lose its success in predicting the shift in the
correlation function.  Simulations show no correlation between spin and
environment (e.g., Barnes \& Efstathiou 1987).
Therefore there should be no shift in the correlation function with
surface brightness as is observed.  This is as much a problem for the SH
picture as the TF relation is for DD.  We might or might not
be able to fix it (cf.\ Mihos et al.\ 1997; Moore et al.\ 1999), but whatever
we come up with is a patch after and against the original fact.

\subsection{Mass and Light}

There is information in the rotation curve beyond the TF relation.
The shapes of rotation curves are also related to luminosity, as noted by
Rubin et al.\ (1985) and Persic \& Salucci (1991).  Though I would not
claim as strict a relation as implied by the `universal rotation curve' of
Persic \& Salucci, a correlation does exist and provides an additional test
(Figure 2).

\begin{figure}[t]
\plotone{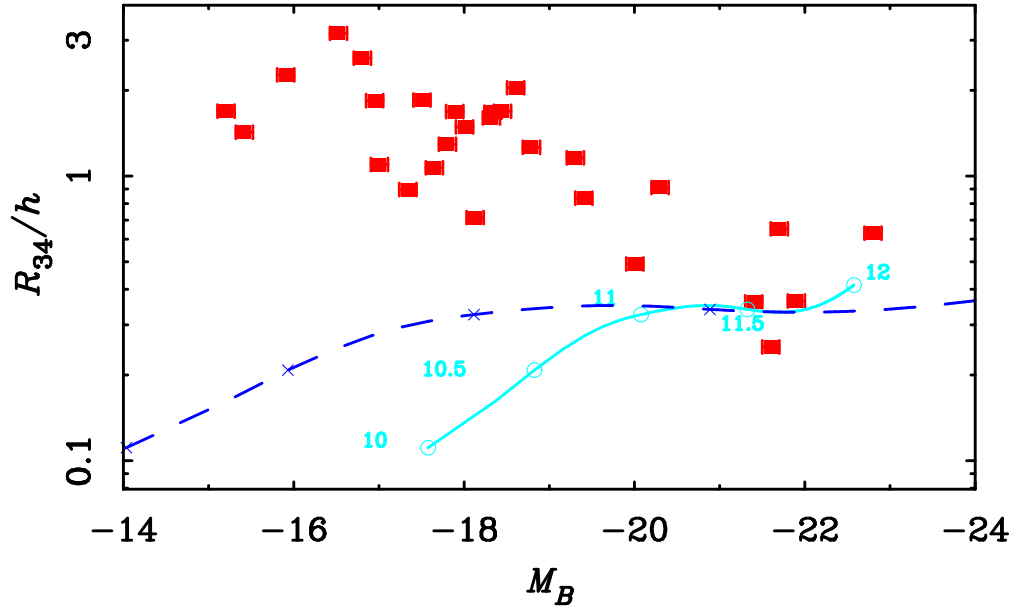}
\caption{The systematic dependence of the shape of rotation curves on
luminosity and disk scale length.  $R_{34}$ is the radius at which $V(R)$
has reached 3/4 of the asymptotic flat velocity.  A large $R_{34}$ corresponds
to a slowly rising rotation curve.  $R_{34}$ varies greatly
between galaxies of the same luminosity but different surface brightness
when $R_{34}$ is measured in kpc.  However, there is a good correlation
in this diagram when $R_{34}$ is normalized by
the disk scale length $h$.  Some, and perhaps most, of the
scatter is attributable to observational uncertainty.  Also shown are
model predictions of Dalcanton et al.\ (1997), with numbers labeling
points by the logarithm of the halo mass.  The models track in the opposite
sense of the data, a problem generic to SH models which have not been tuned
to fit the data.
}
\label{R34}
\end{figure}

LSB galaxies adhere to the relation between luminosity and rotation curve
shape, provided that the radius is measured in units
of the disk scale length (see also Verheijen, these proceedings).
This implies that a good estimate of the rotation curve of any galaxy can
be made from measurements of only two {\it photometric\/} parameters ($L, h$).
Even though the dynamics are dominated by dark matter, we need only know
the distribution of luminous matter to predict the rotation curve.

This strong coupling of mass and light (as long stressed by Sancisi)
is a general problem.
It always leads to fine-tuning paradoxes, with the tail wagging the dog.
The dominant, spherical halo composed of non-baryonic dark matter
simply should not be so intimately related to the details of the luminous disk.

\subsection{CDM and NFW}

I have stressed the importance of {\it a priori\/} predictions.
There has been substantial progress in numerical investigations of the
formation of halos in CDM simulations (e.g., NFW).  These result in an
apparently universal profile describable by two parameters, a concentration
$c$ and a scale $V_{200}$.  These are tightly correlated, effectively forming a
one parameter family for any given cosmology.

The data for LSB galaxies provide a good test of the these predictions
(Figure 3).  Because they have such large mass discrepancies, complications
due to the baryonic component are minimized.  The precise value of
$\Upsilon_*$, adiabatic compression, feedback, etc., simply do not matter.

\begin{figure}[t]
\plotone{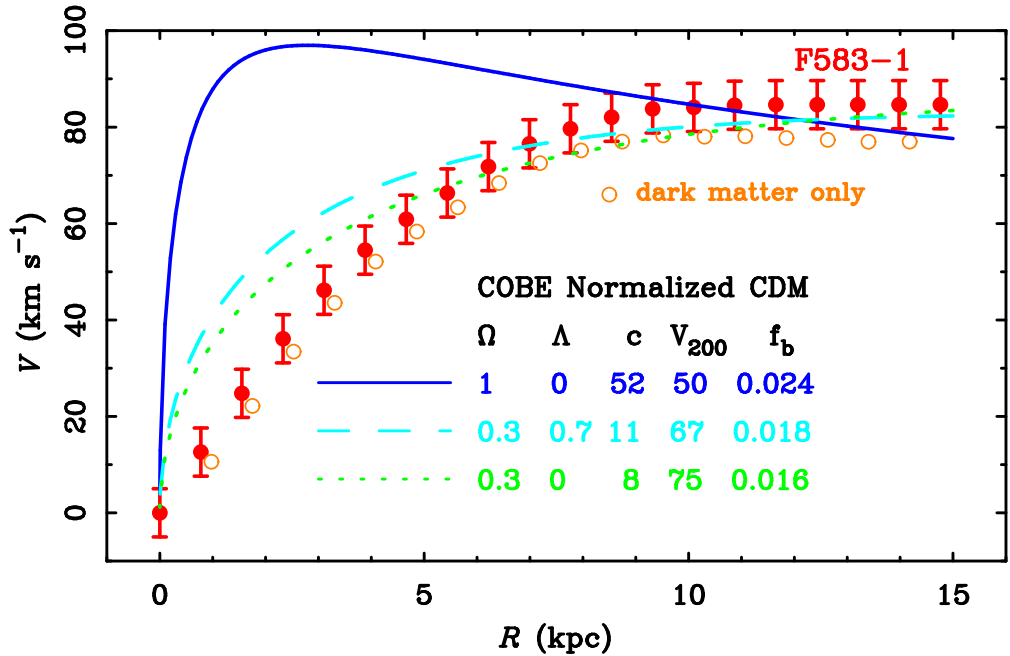}
\caption{The rotation curve of the LSB galaxy F583-1.  The solid
points are the data; the open points are the remainder after the subtraction
of the maximum disk component.  This illustrates the large mass discrepancy
typical of LSB galaxies, even at small radii.  Also shown are the
rotation curves due to NFW halos expected in several cosmologies.
These rise too steeply at small radii, and require very low disk-to-total
masses ($f_b$) in order to match the velocity at large radii.
}
\label{NFW}
\end{figure}

The predicted shape of the inner profile of the dark matter halo goes as
$\rho \propto r^{-\gamma}$ with $\gamma = 1$ according to NFW.  Moreover,
the halo parameters are specified by the cosmology.  Figure 3
shows, in several cosmologies,  the expected rotation curves for NFW halos
chosen to be a close match to the illustrated galaxy.  The predictions fail
in two ways:  NFW halos have rotation curves which rise too steeply,
and require very low disk-to-total mass ratios.

The first failing is obvious by inspection, as first noted by
Flores \& Primack (1994) and Moore (1994).  The data for LSB galaxies
confirm and extend the conclusions of these works.  There have been various
attempts to wiggle out of this problem, without success
(McGaugh \& de Blok 1998a).  Adiabatic contraction should
occur at some level, but acts in the wrong direction.
One might suppose instead that dwarf and LSB galaxies suffered massive
baryonic blow outs following an episode of intense star formation
which rearrange the dark matter in the required way
(Navarro et al.\ 1996).  There is no empirical evidence that this ever
happened.  Gas has not been blown away (McGaugh \& de Blok 1997), nor
is there any debris around LSB galaxies (Bothun et al.\ 1994).
It now appears that the idea is unworkable even in theory
(MacLow \& Ferrara 1998).

\begin{figure}[t]
\plotone{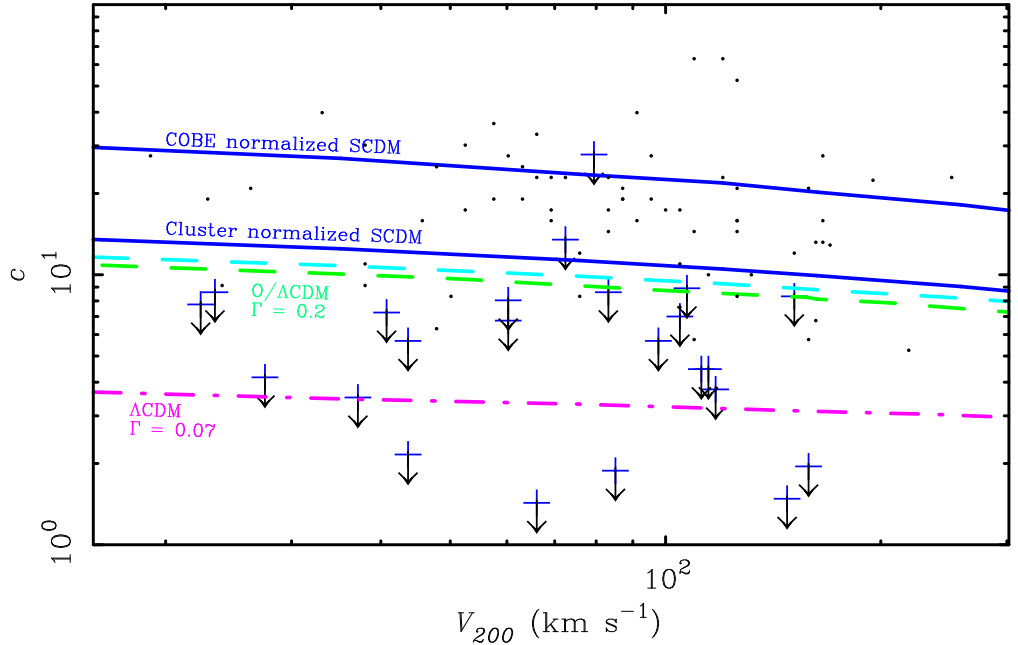}
\caption{Predicted and measured NFW halo parameters $(c,V_{200})$.
These are expected to be tightly correlated in any given cosmology (lines).
Different cosmologies are represented by different lines, as labeled.
The data place upper limits on $c$ by assuming stars have no mass
(Pickering et al.\ 1997).
This is unrealistic in HSB galaxies, leading to uninteresting limits
(dots).  The stellar mass is negligible in LSB galaxies, which place
strong upper limits (crosses with arrows) on $c$.
No plausible cosmology produces sufficiently low concentrations
$c$ to satisfy the data, even with the favorable assumption of low $H_0$.
}
\label{NFWcV}
\end{figure}

The problem is soluble if the amount of dark matter in the inner parts
of the initial halos can be made low enough.  In the context of NFW halos,
this means very low $c$.  This in turn requires a very contrived cosmology
(Figure 4).  Standard ($\Omega = 1$) CDM produces halos which are much too
concentrated.  Low $\Omega$ models (with or without a cosmological constant)
fare better, but still produce halos in which $c$ significantly exceeds the
upper limits imposed by the LSB galaxy data.  Using Navarros's code, the only
way I have found to lower $c$ further is to either reduce the normalization
of the power spectrum well below the values inferred from COBE or clusters
of galaxies, or to choose cosmological parameters ($\Omega, H_0$) which
yield a shape parameter $\Gamma < 0.1$.  This is inconsistent with the
large scale structure constraint $0.2 < \Gamma < 0.3$ (Peacock \& Dodds 1994).

This is very bad for CDM, or at least the NFW realization thereof.
Could something be wrong with the profile suggested by these simulations?
Initially, there seemed to be a good consensus between different simulations.
More recently, this has become rather controversial.
Most workers give $\gamma = 1$ or $\gamma > 1$
(Dubinski 1994; Cole \& Lacey 1996; Moore et al.\ 1998;
Tissera \& Dominguez-Tenreiro 1998; Nusser \& Sheth 1998).
Any $\gamma \ge 1$ is inconsistent with the data.

On the other hand, Kratzsov et al.\ (1998) give $\gamma \approx 0.2$.  This is
inconsistent with other simulations but consistent with the data (see Primack,
these proceedings).  Indeed, Kratzsov et al.\ (1998) constrain $\gamma$
by fitting the data for dwarf and LSB galaxies.  This constrained
profile may be consistent with their simulations, but it hardly constitutes
an independent test.

\begin{figure}[t]
\plotone{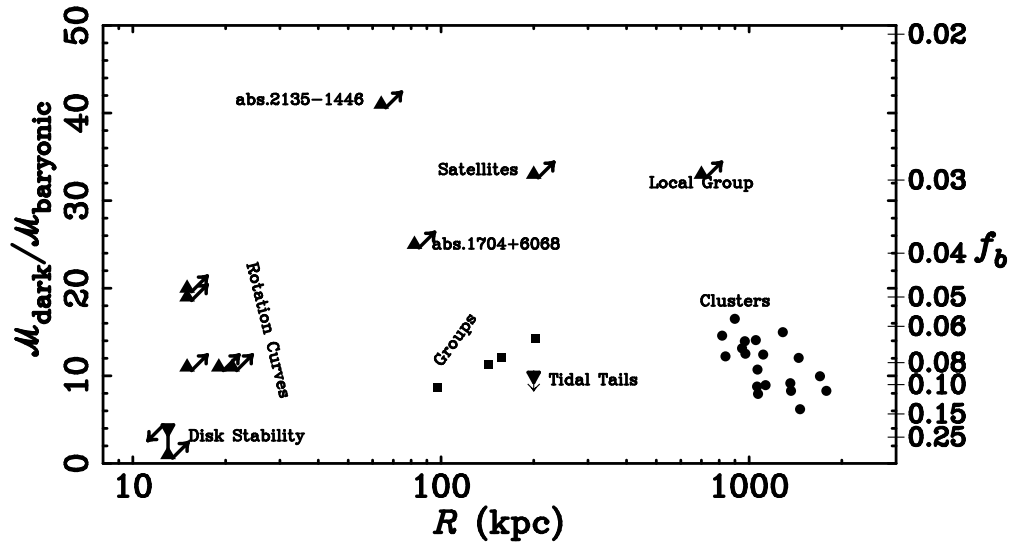}
\caption{Measurements of and limits on the baryon fraction in various systems
(McGaugh \& de Blok 1998a).
A single universal baryon fraction is not evident.  Indeed, a number of
limits are in conflict.  For example, the limits for LSB galaxies refer only
to the edge of the observed rotation curve (de Blok \& McGaugh 1997).
If we analyze them in the same
way as clusters, the limits at lower left become points in the upper left
portion of the diagram.
}
\label{MdMbRh}
\end{figure}

There is another problem in the baryon fraction.
The general picture is of a universal baryon fraction $f_b$ which is
constant from halo to halo (giving rise, among other things, to the
TF relation).
A great deal of emphasis has been placed on measurements of $f_b$
in clusters of galaxies (e.g., Evrard 1997).  However, the same exercise
can be done in other systems as well, with the result that there is no
indication of a universal baryon fraction (Figure 5).
Indeed, a number of limits are in blatant contradiction.
There are various outs, too numerous to discuss here
(see McGaugh \& de Blok 1998a).  To reconcile the cluster value of
$f_b \approx 0.1$ with the LSB galaxy value of $f_b \approx 0.02$
suggested in Figure 3, one might suggest that not all of the baryons have
been incorporated into the disk.
If so, there is no reason to think the fraction which have been will be
the same for all galaxies.  Yet it must be.
Any scatter would propagate into the TF relation,
for which there is no room in the error budget.

I have long been a believer in the CDM paradigm.
Yet I am forced to conclude that the accumulated evidence now weighs
heavily against it.  Unfortunately, the existence of dark matter is not
explicitly falsifiable. Theories in which the dominant mass component is
invisible have sufficient flexibility to accommodate anything.

\section{A Physical Scale in the Mass Discrepancy}

Galaxy formation theory fails to predict a mass distribution which yields
the correct effective force law.  So
let us take a more empirical approach.  A useful measure of the severity
of the mass discrepancy is the ratio of the required gravitating mass to
the total observed luminous mass.  One would not expect dark matter to be
aware of any particular physical scale.  Is there one in the data?

For some reason, our brains think first in terms of linear size.  Galaxies
are big.  Is there anything special about the mass discrepancy at some large
length scale?  The answer is a resounding {\bf no} (Figure 6a).
There are galaxies for which the mass discrepancy is not apparent until
quite large radii, and others (predominantly LSB galaxies) in which it
appears nearly at $R= 0$.  Size does not matter.

\begin{figure}[t]
\plotone{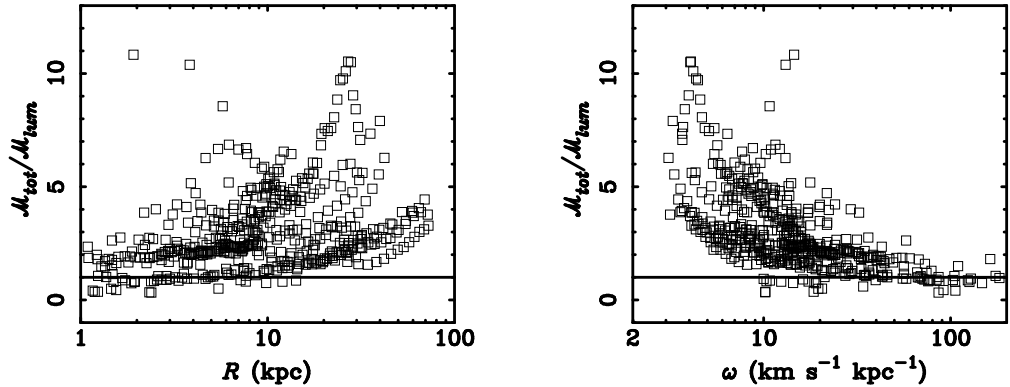}
\caption{The mass discrepancy as a function of (a) radius and (b)
orbital frequency ($\omega = V/R$).  Each point represents one resolved
measurement in the rotation curve of a disk galaxy.  Data for many galaxies
of all luminosities and surface brightnesses are plotted together.
The mass discrepancy is defined as the ratio of dynamical mass
${\cal M}_{tot} = R V^2/G$ to luminous mass 
${\cal M}_{lum} = \Upsilon_* L + {\cal M}_{gas}$.
A stellar mass-to-light ratio $\Upsilon_* = 2$ is assumed, as is
${\cal M}_{gas} = 1.4 {\cal M}_{HI}$.  A line of unity indicates no
mass discrepancy.  There is no preferred scale in either plot.
The mass discrepancy can appear at small as well as large radii (a),
where the eye can perceive some individual rotation curves.  This is
less true in (b), which shows that the mass discrepancy occurs preferentially
at low frequencies, though there is a lot of scatter at a given $\omega$.
Data are taken from the compilation of Sanders (1996) and de Blok \& McGaugh
(1998).
}
\label{MDRWC}
\end{figure}

There are other scales besides linear size.  A plot of the mass
discrepancy against orbital frequency (Figure 6b) begins to show some
organization.  While there is still a lot of scatter, this is a hint that
there may be some interesting scale.

\begin{figure}
\plotone{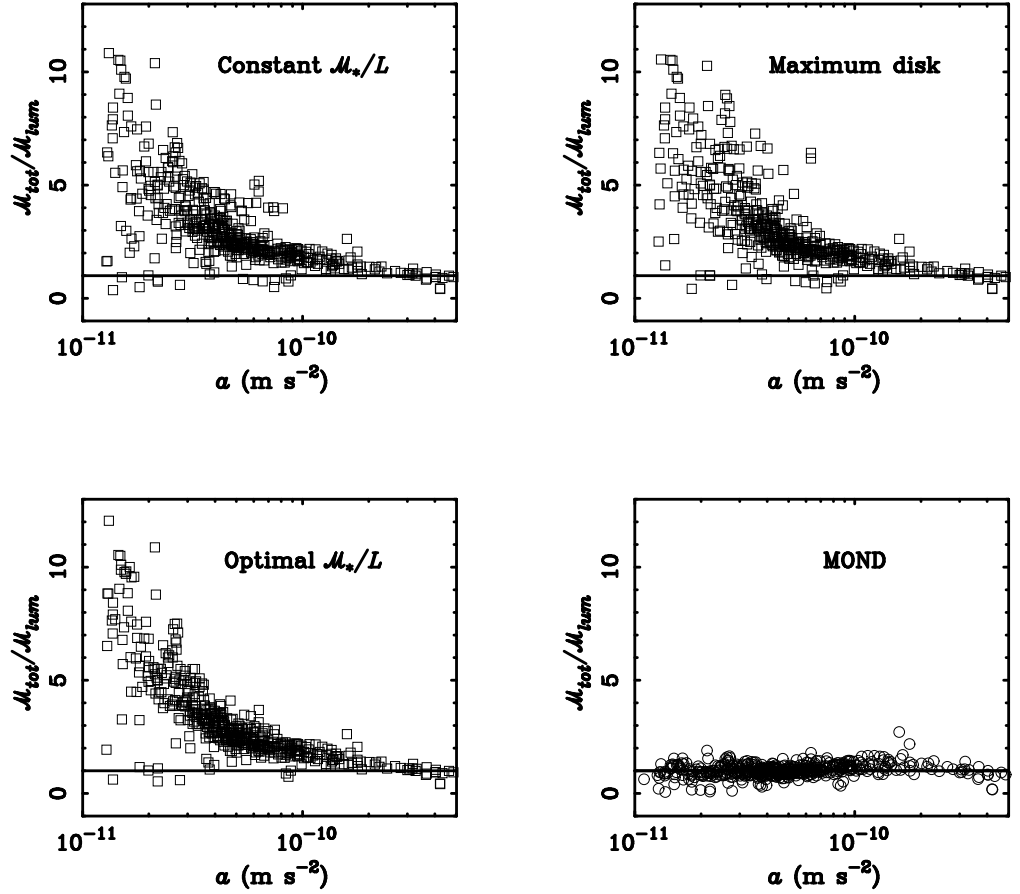}
\caption{Like the preceding Figure, but now with the abscissa in terms
of centripetal acceleration $V^2/R$.  The data all fall
together:  it is no longer possible to distinguish individual rotation curves.
The mass discrepancy in spiral galaxies occurs uniformly at a particular
acceleration scale of $\sim 10^{-10}\>{\rm m}\>{\rm s}^{-2}$.
This occurs regardless of what we assume for the mass-to-light ratio,
as the mass discrepancy is large in LSB galaxies even in the maximum disk
case (b).  The only way to avoid the appearance of a particular acceleration
scale is to allow $\Upsilon_* \rightarrow 0$, in which case there is a mass
discrepancy everywhere.  Moreover, it is possible to choose an optimal
mass-to-light ratio which minimizes the scatter (c) as far as possible
given experimental uncertainty.  The value of each galaxy's $\Upsilon_*$
in (c) could
be chosen to minimize the scatter in this plot, but this is not what has been
done.  Instead, a specific, universal prescription is applied: the optimal
value of $\Upsilon_*$ is that indicated by MOND fits to
the rotation curves.  MOND resolves the mass
discrepancy in disk galaxies (d), providing an accurate description of
the effective force law therein.
}
\label{MDAall}
\end{figure}

A scale which is truly unique to galaxies is that of low acceleration.
The centripetal acceleration which keeps stars in their orbits is typically
$\le 1\>{\rm \AA}\,{\rm s}^{-2}$, only one part in $10^{11}$ of what we
experience at the surface of the earth.  The data show an enormous
regularity in terms of acceleration (Figure 7: see also Sanders 1990).
It matters little how
we compute the luminous mass (assuming constant $\Upsilon_*$ or maximum disk).
The reason for this is that LSB galaxies have large mass discrepancies
in either case, and also have the lowest accelerations.

To the best of my knowledge, there is no dark matter based prediction of
this phenomenology in the literature.  Attempts to impose it by fixing
the surface density of dark halos fail (Sanders \& Begeman 1994).
It makes no sense to me that
dark matter should be aware of a particular physical scale in this fashion.
This is reflected in the difficulties discussed in \S 2.

The regularity of the data strongly
suggest that some universal phenomenon is at work.
There is one suggestion that the usual dynamics should be modified at an
acceleration scale: MOND (Milgrom 1983a).
Milgrom (1983b) made a series of predictions specific to
LSB galaxies.  I find this quite remarkable, as at that time
much of the debate over LSB galaxies was about whether they existed at all,
with the majority opinion being negative.  Now we know they exist, and have the
data to test Milgrom's predictions.

Every prediction Milgrom made about LSB galaxies in 1983 is confirmed
(McGaugh \& de Blok 1998b).

There are many claims to have falsified MOND.
An extensive survey of the literature turns up no credible empirical evidence
against Milgrom's hypothesis (McGaugh \& de Blok 1998b).  Indeed, there are
many cases which are cited as evidence against MOND which can equally well
be argued to support it (dwarf Spheroidals being one example).
Are we failing to see the forest for the trees?

It remains an open question whether MOND works in all places where a mass
discrepancy is inferred.  But it is extremely effective at resolving the
mass discrepancy in disk galaxies (Figure 7d).  This is telling us something.
For all practical purposes, MOND is the effective force law in disk galaxies.

Irrespective of whether MOND is correct as a theory, it does
constitute an observed phenomenology.  As such, it provides a very strong
test of dark matter galaxy formation theories.  The requirement is this:
for any model, it must be possible to apply the MOND formula to the
disk component and obtain the rotation curve which is actually produced
by the combination of disk and halo components.

There is very little freedom to achieve this.  All the many
parameters of conventional models must be encapsulated in one
tightly constrained parameter, ${\cal Q}$.  This is the ratio of
the stellar mass-to-light ratio that is required to obtain a MOND
fit to that which is `correct' in the model
(${\cal Q} \equiv \Upsilon_*^{MOND}/\Upsilon_*^{DM}$).  Since the mass-to-light
ratios indicated by MOND are already quite reasonable for stellar populations
(Sanders 1996, de Blok \& McGaugh 1998),
this is a very restrictive requirement indeed: ${\cal Q} \approx 1$
with no other adjustable parameters.
Nevertheless, this is what is required to reproduce the observed phenomenology.

Until there is a credible explanation for the MOND phenomenology in
the framework of the standard paradigm, we should be as skeptical of the
existence of dark matter as we are of the need to modify Newton's Laws.

\acknowledgments  I am grateful to Erwin de Blok for our long and fruitful
collaboration, and to Vera Rubin, Renzo Sancisi, Bob Sanders,
and Jerry Sellwood for many animated discussions of disk dynamics.

\begin{question}{Spergel}
I assert that an acceleration scale does occur naturally in dark matter.
\end{question}
\begin{answer}{McGaugh}
Did you predict this in 1983, or did Milgrom?
\end{answer}


\begin{references}
\reference Barnes, J. \& Efstathiou, G. 1987, \apj, 319, 575
\reference Bothun, G.D., Eriksen, J., \& Schombert, J.M. 1994, \aj, 108, 913
\reference Cole, S. \& Lacey, C. 1996, \mnras, 281, 716
\reference Dalcanton, J. J., Spergel, D. N., \& Summers, F. J. 1997,
        \apj, 482, 659
\reference de Blok, W.J.G., \& McGaugh, S.S. 1997, \mnras, 290, 533
\reference de Blok, W.J.G., \& McGaugh, S.S. 1998, \apj, 508, 132
\reference de Blok, W.J.G., McGaugh, S.S., \& van der Hulst, J.M. 1996, \mnras,
	283, 18
\reference Evrard, A.E. 1997, \mnras, 292, 289
\reference Flores, R.A., \& Primack, J.R. 1994, \apj, 427, L1
\reference Hoffman, G. L., Salpeter, E. E., Farhat, B., Roos, T.,
        Williams, H. \& Helou, G. 1996, \apjs, 105, 269
\reference Kratzsov, A.V., Klypin, A.A., Bullock, J.S., \& Primack, J.R.
	1998, \apj, 502, 48
\reference MacLow, M.-M., \& Ferrara, A. 1998, astro-ph/9801237
\reference Matthews, L.D., van Driel, W., \& Gallagher, J.S. 1998,
	\aj, 116, 2196
\reference McGaugh, S.S. 1992, Ph.D. thesis, University of Michigan
\reference McGaugh, S.S. 1996, \mnras, 280, 337
\reference McGaugh, S.S., \& Bothun, G.D. 1994, \aj, 107, 530
\reference McGaugh, S.S., \& de Blok, W.J.G. 1997, \apj, 481, 689
\reference McGaugh, S.S., \& de Blok, W.J.G. 1998a, \apj, 499, 41
\reference McGaugh, S.S., \& de Blok, W.J.G. 1998b, \apj, 499, 66
\reference Mihos, J.C., McGaugh, S.S., \& de Blok, W.J.G. 1997, \apj, 477, L79
\reference Milgrom, M. 1983a, \apj, 270, 365
\reference Milgrom, M. 1983b, \apj, 270, 371
\reference Mo, H.J., Mao, S., \& White, S.D.M. 1998, \mnras, 295, 319
\reference Mo, H.J., McGaugh, S.S., \& Bothun, G.D. 1994, \mnras, 267, 129
\reference Moore, B. 1994, Nature, 370, 629
\reference Moore, B., Governato, F., Quinn, T., Stadel, J., \& Lake, G.
	1998, \apj, 499, L5
\reference Moore, B., Lake, G., Quinn, T., \& Stadel, J. 1999, \mnras,
	in press
\reference Navarro, J. F., Eke, V. R., \& Frenk, C. S. 1996,
        \mnras, 283, L72
\reference Navarro, J.F., Frenk, C.S., \& White, S.D.M. 1997, \apj, 490,
	493 (NFW)
\reference Nusser, A., \& Sheth, R. 1998, astro-ph/9803281
\reference Peacock, J.A., \& Dodds, S.J. 1994, \mnras, 267, 1020
\reference Persic, M. \& Salucci, P. 1991, \apj, 368, 60
\reference Pickering, T.E., Impey, C.D., van Gorkom, J.H., \& Bothun, G.D.
	1997, \aj, 114, 1858
\reference Rubin, V.C, Burstein, D., Ford, W.K., \& Thonnard, N.  1985,
	\apj, 289, 81
\reference Sanders, R.H. 1990, \aapr, 2, 1
\reference Sanders, R.H. 1996, \apj, 473, 117
\reference Sanders, R.H., \& Begeman, K.G. 1994, \mnras, 266, 360
\reference Sprayberry, D., Bernstein, G.M., Impey, C.D.,
        \& Bothun, G.D. 1995b, ApJ, 438, 72
\reference Tissera, P. \& Dominguez-Tenreiro, R. 1998, \mnras, 297, 177
\reference Tully, R.B., \& Verheijen, M.A.W. 1997, \apj, 484, 145
\reference Zwaan, M.A., van der Hulst, J.M., de Blok, W.J.G., \&
        McGaugh, S.S. 1995, \mnras, 273, L35
\end{references}
\end{document}